\documentclass[10pt, conference, compsocconf,letterpaper,twocolumn]{IEEEtran}

\usepackage[rgb]{xcolor}
\usepackage{graphicx}
\usepackage{paralist}
\usepackage{url}
\usepackage{subfig}
\usepackage{adjustbox}
\usepackage{tikz}
\usepackage{scalefnt}
\usepackage{pgf}

\usepackage{todonotes}
\usepackage[]{algorithm2e}

\usetikzlibrary{arrows}
\usetikzlibrary{patterns,calc,decorations.markings}
\usetikzlibrary{positioning}
\usetikzlibrary{calc,shapes,arrows,backgrounds,positioning,patterns}
\usetikzlibrary{shapes.misc}
\usetikzlibrary{decorations.pathmorphing}
\usetikzlibrary{decorations.pathreplacing}
\usetikzlibrary{decorations.shapes}
\usetikzlibrary{decorations.text}
\usetikzlibrary{decorations.markings}
\usetikzlibrary{decorations.fractals}
\usetikzlibrary{decorations.footprints}

\makeatletter
\def\convertto#1#2{\strip@pt\dimexpr #2*65536/\number\dimexpr 1#1}
\makeatother



\tikzset{
  big arrow/.style={
    decoration={markings,mark=at position 0.9999 with {\arrow[scale=1.5,#1]{latex}}},
    postaction={decorate},
    shorten >=1.2pt
},
  big arrow/.default=blue
}

\tikzstyle{block} = [rectangle,fill=none,solid, draw=black, text centered,minimum width=2em, minimum height=2em,inner sep=0pt]
\newcommand*\circled[1]{\tikz[baseline=(char.base)]{ \node[shape=circle, fill=white,draw,inner sep=0pt,text width= 1 em ,minimum size = 1 em, text centered] (char) {#1};}}
\tikzstyle{treenode} = [rectangle,fill=white,solid, draw=black, text centered,minimum size=1.5em, text width=1.5em,inner sep=0pt]
\DeclareRobustCommand\supernode[1]{\tikz[baseline=(char.base)]{ \node[shape=rectangle, fill=white,draw,inner sep=0pt,text width= 1 em ,minimum size = 1 em, text centered] (char) {#1};}}

\DeclareRobustCommand\fdots[0]{\tikz[baseline=(char.south)]{ 
\node[shape=circle,inner sep=0pt,minimum size=.8\baselineskip] (char){}; \node[shape=circle, fill=black,inner sep=0pt,minimum size = 0.4 em] {};}}

\newcommand{\elm}[2]{\fill[fill=black] ($(#1 em,-#2 em )+(0.5 em,-0.5 em)$) circle [radius=0.4 em]}

\newcommand{\edblk}[3]{
\node[draw=none,minimum width=#2 em, minimum height=#2 em] at ($(#3 em,-#3 em) + 0.5*(0,6 em)$) {\scalefont{1.5}#1};
}

\newcommand{\dblk}[3]{
    \pgfmathtruncatemacro{\start}{#3 -0.5*#2};
    \pgfmathtruncatemacro{\stop}{#3 +0.5*#2-1};
  \foreach \i in {\start,...,\stop}{
    \foreach \j in {\i,...,\stop}{
      \elm{\i}{\j};
    }
  }
\edblk{\supernode{#1}}{#2}{#3};
}

\newcommand{\mapblock}[4]{
    \pgfmathtruncatemacro{\curproc}{ int(mod(#2-1,#3))*#4 + int(mod(#1-1,#4)) +1 };
}

\newcommand{\colorproc}[3]{
  \pgfmathsetmacro{\h}{#1 /(#2*#3)}
  \definecolor{proccolor}{hsb}{\h, 0.5, 0.8}
}

\newcommand{\elublk}[5]{
\node[block,draw=none,minimum width=#2em, minimum height=#3em] at (#4em,-#5em) {#1};
}

\newcommand{\lublk}[5]{
    \pgfmathtruncatemacro{\starti}{#4 -0.5*#2};
    \pgfmathtruncatemacro{\stopi}{#4 +0.5*#2-1};
    \pgfmathtruncatemacro{\startj}{#5 -0.5*#3};
    \pgfmathtruncatemacro{\stopj}{#5 +0.5*#3-1};
  \foreach \i in {\starti,...,\stopi}{
    \foreach \j in {\startj,...,\stopj}{
      \elm{\i}{\j};
    }
  }
\elublk{#1}{#2}{#3}{#4}{#5};
}

\tikzset{cross/.style={cross out, draw,ultra thick,rotate=45, 
         minimum size=2*(#1), 
         inner sep=0pt, outer sep=0pt},
cross/.default={0.4em}}

\newcommand{\fillElm}[2]{
\draw[ultra thick] ($(#1 em,-#2 em )+(0.15 em,-0.15 em)+(0,-0.35em)$)
--($(#1 em,-#2 em )+(0.15 em,-0.15 em)+(0.7em,-0.35em)$);
\draw[ultra thick] ($(#1 em,-#2 em )+(0.15 em,-0.15 em)+(0.35em,0)$)
--($(#1 em,-#2 em )+(0.15 em,-0.15 em)+(0.35em,-0.7em)$);
}
\newcommand{\flublk}[5]{
    \pgfmathtruncatemacro{\starti}{#4 -0.5*#2};
    \pgfmathtruncatemacro{\stopi}{#4 +0.5*#2-1};
    \pgfmathtruncatemacro{\startj}{#5 -0.5*#3};
    \pgfmathtruncatemacro{\stopj}{#5 +0.5*#3-1};
  \foreach \i in {\starti,...,\stopi}{
    \foreach \j in {\startj,...,\stopj}{
      \fillElm{\i}{\j}
    }
  }
\elublk{#1}{#2}{#3}{#4}{#5};
}

\newcommand{\ignore}[1]{}

\newcommand{\mumps}{\texttt{MUMPS}\xspace}
\newcommand{\pastix}{\texttt{PASTIX}\xspace}
\newcommand{\superlu}{\texttt{SuperLU\_DIST}\xspace}

\newcommand{\scotch}{\texttt{Scotch}\xspace}

\newcommand{\proc}[1]{\ensuremath{p_{#1}}\xspace}

\newcommand{\psource}[0]{\ensuremath{\proc{\textnormal{source}}}\xspace}
\newcommand{\ptarget}[0]{\ensuremath{\proc{\textnormal{target}}}\xspace}

\urldef{\mailsa}\path|{mjacquelin,egng,kayelick,yzheng}@lbl.gov|    

\newcommand{\REV}[1]{#1}
\newcommand{\M}[1]{\mathbf{\MakeUppercase{#1}}}

\DeclareRobustCommand\circled[1]{\tikz[baseline=(char.base)]{ \node[shape=circle, fill=white,draw,inner sep=0pt,text width= 1 em ,minimum size = 1 em, text centered] (char) {#1};}}

\newcommand{\sympack}[0]{\texttt{symPACK}\xspace}
\newcommand{\upcxx}[0]{\texttt{UPC++}\xspace}
\newcommand{\gasnet}[0]{\texttt{GASNet}\xspace}

\newcommand{\ltq}[0]{\texttt{LTQ}\xspace}
\newcommand{\rtq}[0]{\texttt{RTQ}\xspace}

\newcommand{\fanin}[0]{\textit{fan-in}\xspace}
\newcommand{\fanout}[0]{\textit{fan-out}\xspace}
\newcommand{\fanboth}[0]{\textit{fan-both}\xspace}

\newcommand{\Fanin}[0]{\textit{Fan-in}\xspace}
\newcommand{\Fanout}[0]{\textit{Fan-out}\xspace}
\newcommand{\Fanboth}[0]{\textit{Fan-both}\xspace}

\newcommand{\leftlooking}[0]{\textit{left-looking}\xspace}
\newcommand{\rightlooking}[0]{\textit{right-looking}\xspace}
\begin{document}

\title{An Asynchronous Task-based Fan-Both Sparse Cholesky Solver}

%
%

\author{
  \IEEEauthorblockN{Mathias Jacquelin, Yili Zheng\IEEEauthorrefmark{1}, Esmond Ng, Katherine Yelick}
\IEEEauthorblockA{Lawrence Berkeley National Laboratory\\
\url{mjacquelin@lbl.gov}, \url{yilizheng@google.com}, \url{egng@lbl.gov}, \url{kayelick@lbl.gov}\\
\IEEEauthorrefmark{1}Yili Zheng has since joined Google}
}

\maketitle

\begin{abstract}
Systems of linear equations arise at the heart of many scientific and engineering applications.  Many of these linear systems are sparse; i.e., most of the elements in the coefficient matrix are zero.  Direct methods based on matrix factorizations are sometimes needed to ensure accurate solutions.  For example, accurate solution of sparse linear systems is needed in shift-invert Lanczos to compute interior eigenvalues.  The performance and resource usage of sparse matrix factorizations are critical to time-to-solution and maximum problem size solvable on a given platform.

In many applications, the coefficient matrices are symmetric, and exploiting symmetry will reduce both the amount of work and storage cost required for factorization. 
When the factorization is performed on large-scale distributed memory platforms, communication cost is critical to the performance of the algorithm. At the same time, network topologies have become increasingly complex, so that modern platforms exhibit a high level of performance variability.  This makes scheduling of computations an intricate and performance-critical task.

In this paper, we investigate the use of an asynchronous task paradigm, one-sided communication and dynamic scheduling in implementing sparse Cholesky factorization (\sympack) on large-scale distributed memory platforms. 
Our solver \sympack relies on efficient and flexible communication primitives provided by the \upcxx
library.
Performance evaluation shows good scalability and that \sympack outperforms 
state-of-the-art parallel distributed memory factorization packages,
validating our approach on practical cases.

\end{abstract}


\section{Introduction}


Symmetric positive definite systems of linear equations arise in the solution of many scientific and engineering problems.
Efficient solution of such linear system is important for the overall performance of the application codes.
In this paper, we consider direct methods for solving a sparse symmetric positive definite linear system, which are based on Cholesky factorization.  While direct methods can be expensive for large matrices, in terms of execution times and storage requirement when compared to iterative methods, they have the advantage that they terminate in a finite number of operations.  Also, direct methods can handle linear systems that are ill conditioned or the situation when there are many multiple right-hand sides.  An example of ill-conditioned linear systems is in the computation of interior eigenvalues of a matrix using the shift-invert Lanczos algorithm.

We propose a new implementation of sparse Cholesky factorization using an \emph{asynchronous task paradigm}.  We introduce a parallel distributed memory solver called \sympack. By using a task-based 
formalism and dynamic scheduling techniques within a
node, \sympack achieves good strong scaling on modern supercomputers.

An outline of the paper is as follows.  In Section~\ref{background}, we provide some background on sparse Cholesky factorization.  In Section~\ref{sympack}, we present our implementation in \sympack.  The asynchronous paradigm is described in Section~\ref{asynchronous}.  Some numerical results are presented in Section~\ref{performance}, followed by some concluding remarks in Section~\ref{conclusion}



\section{Background on Cholesky factorization}\label{background}

In the following, we give some background on 
Cholesky factorization and how symmetry and sparsity can be taken into account. We first review the basic Cholesky algorithm for dense matrices
and then detail how it can be modified to
handle sparse matrices efficiently. We also present fundamental
notions on sparse matrix computations before reviewing the work
related to sparse Cholesky factorization.



\subsection{The basic algorithms}

Let $\M{A} = [ a_{i,j} ]$ be an $n$-by-$n$ symmetric positive definite matrix. The Cholesky
algorithm factors the matrix $\M{A}$ into
\begin{equation}
  \M{A} = \M{L}\M{L}^T,
\end{equation}
where $\M{L} = [ \ell_{i,j} ]$ is a lower triangular matrix, and $\M{L}^T$ is the transpose of $\M{L}$ and is upper triangular.  The factorization thus allows symmetry to be exploited, since only $\M{L}$ needs to be computed and saved.

\begin{algorithm}
\DontPrintSemicolon
\For{column $j = 1$ to $n$}{
  $\ell_{j,j} = \sqrt{a_{j,j}}$\;
  \For{row $i=j+1$ to $n$}{
    $\ell_{i,j} = a_{i,j} / \ell_{j,j}$\; 
  } 
  \;
  \For{column $k=j+1$ to $n$}{
    \For{row $i=k$ to $n$}{
      $a_{i,k} = a_{i,k} - \ell_{i,j} \cdot \ell_{k,j}$ \;
    }
  }

}
\caption{Basic Cholesky algorithm}
\label{alg.chol}
\end{algorithm}

The basic Cholesky factorization algorithm, given in Alg.~\ref{alg.chol}, can be
described as follows:
\begin{compactenum}
\item Current column $j$ of $\M{L}$ is computed using column $j$ of $\M{A}$.
\item Column $j$ of $\M{L}$ is used to update the remaining columns of $\M{A}$.
\end{compactenum}
If $\M{A}$ is a dense matrix, then every column $k$, $k>j$, is updated.

Once the factorization is computed, the solution to the original linear system can be obtained by solving two triangular linear systems using the Cholesky factor $\M{L}$.


\subsection{Cholesky factorization of sparse matrices}

For large-scale applications, $\M{A}$ is often {\em sparse\/}, meaning that most of the elements of $\M{A}$ are zero.  When the Cholesky factorization of $\M{A}$ is computed, some of the zero entries will turn into nonzero (due to the subtraction operations in the column updates; see Alg.~\ref{alg.chol}).  The extra nonzero entries are referred to as {\em fill-in\/}.  For in-depth discussion of sparse Cholesky factorization, the reader is referred to~\cite{GeorgeLiu81}.

Following is an important observation in sparse Cholesky factorization.  
It is expected that the columns of $\M{L}$ will become denser and denser as one moves from the left to the right.  This is due to the fact that the fill-in in one column will result in additional fill-in in subsequent columns.  Thus, it is not uncommon to find groups of consecutive columns that eventually share essentially the same zero-nonzero structure.  Such a group of columns is referred to as a {\em supernode\/}.  To be specific, if columns $i$, $i+1$, $\cdots$, $j$ form a supernode, then the diagonal block of these columns will be completely dense, and row $k$, $j+1 \leq k \leq n$, within the supernode is either entirely zero or entirely nonzero.

\begin{figure}[htbp]
\centering
\subfloat[][Structure of Cholesky factor $\M{L}$]{
\label{fig.cholesky_factor}
\begin{adjustbox}{width=.50\linewidth}
\begin{tikzpicture}

    \pgfmathtruncatemacro{\Pr}{1};
    \pgfmathtruncatemacro{\Pc}{4};

\mapblock{0}{0}{\Pr}{\Pc} \colorproc{\curproc}{\Pr}{\Pc}
\draw[fill=proccolor!30,draw=none] (0em,0em) rectangle (2em ,-30.5em);
\mapblock{1}{0}{\Pr}{\Pc} \colorproc{\curproc}{\Pr}{\Pc}
\draw[fill=proccolor!30,draw=none] (2em,-2em) rectangle (4em ,-30.5em);
\mapblock{2}{0}{\Pr}{\Pc} \colorproc{\curproc}{\Pr}{\Pc}
\draw[fill=proccolor!30,draw=none] (4em,-4em) rectangle (8em ,-30.5em);
\mapblock{3}{0}{\Pr}{\Pc} \colorproc{\curproc}{\Pr}{\Pc}
\draw[fill=proccolor!30,draw=none] (8em,-8em) rectangle (10em ,-30.5em);
\mapblock{4}{0}{\Pr}{\Pc} \colorproc{\curproc}{\Pr}{\Pc}
\draw[fill=proccolor!30,draw=none] (10em,-10em) rectangle (16em ,-30.5em);
\mapblock{5}{0}{\Pr}{\Pc} \colorproc{\curproc}{\Pr}{\Pc}
\draw[fill=proccolor!30,draw=none] (16em,-16em) rectangle (18em ,-30.5em);
\mapblock{6}{0}{\Pr}{\Pc} \colorproc{\curproc}{\Pr}{\Pc}
\draw[fill=proccolor!30,draw=none] (18em,-18em) rectangle (20em ,-30.5em);
\mapblock{7}{0}{\Pr}{\Pc} \colorproc{\curproc}{\Pr}{\Pc}
\draw[fill=proccolor!30,draw=none] (20em,-20em) rectangle (23em ,-30.5em);
\mapblock{8}{0}{\Pr}{\Pc} \colorproc{\curproc}{\Pr}{\Pc}
\draw[fill=proccolor!30,draw=none] (23em,-23em) rectangle (25em ,-30.5em);
\mapblock{9}{0}{\Pr}{\Pc} \colorproc{\curproc}{\Pr}{\Pc}
\draw[fill=proccolor!30,draw=none] (25em,-25em) rectangle (30em ,-30.5em);

\node at (19.5em,1em) {\scalefont{2}Processor list:};
\mapblock{0}{0}{\Pr}{\Pc} \colorproc{\curproc}{\Pr}{\Pc}
\node[treenode,minimum size=3em, draw=black, fill=proccolor!30,ultra thick] at (15em,-1.5em){\scalefont{2}$\proc{0}$};
\mapblock{1}{0}{\Pr}{\Pc} \colorproc{\curproc}{\Pr}{\Pc}
\node[treenode,minimum size=3em, draw=black, fill=proccolor!30,ultra thick] at (18em,-1.5em){\scalefont{2}$\proc{1}$};
\mapblock{2}{0}{\Pr}{\Pc} \colorproc{\curproc}{\Pr}{\Pc}
\node[treenode,minimum size=3em, draw=black, fill=proccolor!30,ultra thick] at (21em,-1.5em){\scalefont{2}$\proc{2}$};
\mapblock{3}{0}{\Pr}{\Pc} \colorproc{\curproc}{\Pr}{\Pc}
\node[treenode,minimum size=3em, draw=black, fill=proccolor!30,ultra thick] at (24em,-1.5em){\scalefont{2}$\proc{3}$};



\dblk{1}{2}{1};
\dblk{2}{2}{3};
\dblk{3}{4}{6};
\dblk{4}{2}{9};
\dblk{5}{6}{13};
\dblk{6}{2}{17};
\dblk{7}{2}{19};
\dblk{8}{3}{21.5};
\dblk{9}{2}{24};
\dblk{10}{5}{27.5};

\lublk{}{2}{2}{1}{3};
\lublk{}{2}{1}{3}{12};
\lublk{}{2}{3}{3}{15};

\lublk{}{4}{2}{6}{9};
\lublk{}{2}{6}{9}{13};


%
\lublk{}{6}{2}{13}{24};
%

\lublk{}{2}{3}{17}{21.5};
\lublk{}{2}{2}{17}{26};
\lublk{}{2}{1}{17}{29};

\lublk{}{2}{3}{19}{21.5};

\lublk{}{3}{2}{21.5}{24};
\flublk{ }{3}{2}{21.5}{26};
\flublk{ }{3}{1}{21.5}{29};

\lublk{}{2}{4}{24}{27};

%
%
%



\draw (0em,0em) --  (0em ,-30em);
\draw (2em ,0em) -- (2em ,-30em);
\draw (4em ,-2em) -- (4em ,-30em);
\draw[] (8em ,-4em) -- (8em ,-30em);
\draw (10em,-8em) -- (10em,-30em);
\draw (16em,-10em) -- (16em,-30em);
\draw[] (18em,-16em) -- (18em,-30em);
\draw (20em,-18em) -- (20em,-30em);
\draw (23em,-20em) -- (23em,-30em);
\draw[] (25em,-23em) -- (25em,-30em);
\draw (30em,-25em) -- (30em,-30em);

\end{tikzpicture}
\end{adjustbox}
}
\subfloat[Supernodal elimination tree of matrix $\M{A}$]{
\label{fig.supernodal_etree}
\begin{adjustbox}{width=.45\linewidth}
\begin{tikzpicture}

    \pgfmathtruncatemacro{\Pr}{1};
    \pgfmathtruncatemacro{\Pc}{4};

\clip (-9em,-2em) rectangle +(18em,-19em);

\node[treenode] (10) at (0,-4em) {10};
\node[treenode] (9) at (0,-8em) {9};

\node[treenode] (8) at (-4 em,-12em) {8};
\node[treenode] (5) at (4 em,-12em) {5};

\node[treenode] (6) at (-7 em,-16em) {6};
\node[treenode] (7) at (-1 em,-16em) {7};
\node[treenode] (2) at (1 em,-16em) {2};
\node[treenode] (4) at (7 em,-16em) {4};

\node[treenode] (1) at (1 em,-20em) {1};
\node[treenode] (3) at (7 em,-20em) {3};

\draw (10) -- (9);
\draw (9) -- (5);
\draw (9) -- (8);
\draw (5) -- (2);
\draw (5) -- (4);
\draw (8) -- (6);
\draw (8) -- (7);

\draw (2) -- (1);
\draw (4) -- (3);

\end{tikzpicture}
\end{adjustbox}
}
\caption{Sparse matrix $\M{A}$ partitioned into supernodes, \supernode{i} denotes the $i$-th supernode.
\fdots{} represents original nonzero elements in $\M{A}$, while \textbf{+} denotes fill-in entries. Colors correspond to the 4 distributed memory nodes on which supernodes are mapped in a 1D-cyclic way.
}
\end{figure}
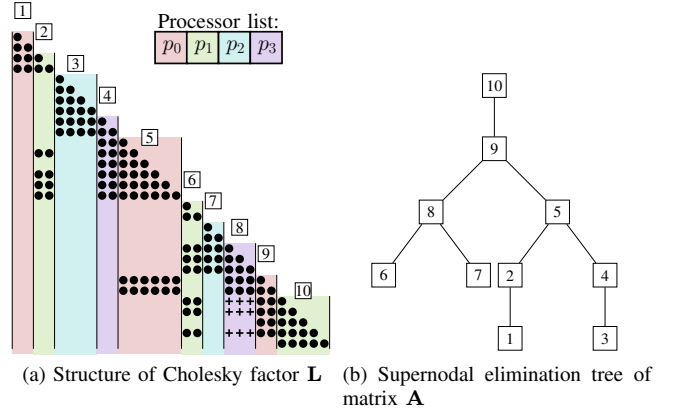

\emph{Fill-in} entries and \emph{supernodes} of a sample symmetric matrix are depicted in Figure~\ref{fig.cholesky_factor}. In this example, 10 supernodes are found. Fill-in entries are created in
supernode 8 because of the nonzero entries in supernode 6.

The {\em elimination tree\/} of $\M{A}$ (or $\M{L}$) is a very important and useful tool in sparse Cholesky factorization.  It is an acyclic graph that has $n$ vertices $\{ v_1 , v_2 , \cdots , v_n \}$, with $v_i$ corresponding to column $i$ of $\M{A}$.  Suppose $i > j$.  There is an edge between $v_i$ and $v_j$ in the elimination tree if and only if $\ell_{ij}$ is the {\em first\/} off-diagonal nonzero entry in column $j$ of $\M{L}$.  Thus, $v_i$ is called the {\em parent\/} of $v_j$ and $v_j$ is a {\em child\/} of $v_i$.  The elimination tree contains a lot of information regarding the sparsity structure of $\M{L}$ and the dependency among the columns of $\M{L}$.  See \cite{Liu90} for details.

An elimination tree can be expressed in terms of supernodes rather than column. In such a case, it is referred to as a \textit{supernodal elimination tree}. An example of such tree is
depicted in Figure~\ref{fig.supernodal_etree}.

\subsection{Scheduling in parallel sparse Cholesky factorization}

In the following, we discuss scheduling of the computation in the numerical factorization.
The only constraints that have to be respected are the numerical dependencies among the columns:
column $k$ of $\M{A}$ has to be updated by column $j$ of $\M{L}$, for any $j < k$ such that $\ell_{k,j} \neq 0$, but
the order in which the updates occur is mathematically irrelevant,
as long as the updates are performed before column $k$ of $\M{A}$ is factored.
There is therefore significant freedom in the scheduling of computational tasks that factorization algorithms
can exploit.

For instance, on sequential platforms, this has led to two well-known variants of the Cholesky
factorization algorithm: \leftlooking and \rightlooking schemes, which 
have been introduced in the context of dense linear 
algebra~\cite{dongarra1987lapack}.
In the \leftlooking algorithm, before column $k$ of $\M{A}$ is factored, all updates coming from columns
$i$ of $\M{L}$ such that $i<k$ and $\ell_{k,i} \neq 0$ are first applied. In that sense, the algorithm is ``looking to the left'' 
of column $k$.
In \rightlooking, after a column $k$ has been factored, every column $i$ such that $k<i$ and $\ell_{i,k} \neq 0$ is
updated by column $k$. The algorithm thus ``looks to the right'' of column $k$.

Distributed memory platforms add the question of where the computations are going to be performed. Various parallel algorithms have been proposed in the literature for Cholesky factorization, such as \mumps~\cite{mumps}, which is based on the multifrontal approach (a variant of right-looking), and \pastix~\cite{pastix}, which is left-looking.

In~\cite{ashcraft-dissertation}, the author classifies parallel Cholesky algorithms into three families: \fanin,
\fanout and \fanboth.

The \fanin family includes all algorithms such that all updates from a column
$k$ to other columns $i$, for $k<i$ such that $\ell_{i,k} \neq 0$, are computed on the processor owning column $k$. When one of these columns, say $i$,
will be factored, the processor owning $i$ will have to ``fan-in'' (or collect) updates from previous columns.

The \fanout family includes algorithms that compute updates from column $k$ to columns $i$, for $k<i$ such that $\ell_{k,i} \neq 0$, on processors
owning columns $i$.
This means that the processor owning column $k$ has to ``fan-out'' (or broadcast) column $k$ of the Cholesky factor.

The \fanboth family generalizes these two families to allow these updates to be performed
on any processors. This family relies on \textit{computation maps} to map computations to processors. 

In the rest of the paper, we will use the term \fanboth \textit{algorithm} as a
shorthand to refer to \textit{an algorithm belonging to the \fanboth family} 
(and similarly for \fanin and \fanout).

\section{A versatile parallel sparse Cholesky algorithm}\label{sympack}

As mentioned in the previous section, there are many ways to schedule the computations as long as the precedence constraints are satisfied.
The Cholesky factorization in \sympack is inspired by the \fanboth
algorithm. This leads to a high level of
versatility and modularity, which allows \sympack to adapt to various
platforms and network topologies.

\subsection{Task-based formulation}

Both \fanboth and \sympack involve three types of operations: 
\textit{factorization, update, aggregation}.
We let $\M{A}$ be an $n$-by-$n$ matrix, and denote these tasks using the following notation\footnote{We use \texttt{MATLAB} notation in this paper.}:
\begin{itemize}
\item \textit{Factorization} $F_{i,i}$: compute column $i$ of the Cholesky factor.
\item \textit{Update} $U_{i,j}$: compute the update from $\ell_{j:n,i}$ to 
	column $j$, with $i < j$ such that $\ell_{j,i} \neq 0$, and put it to an \emph{aggregate vector} $t^{i}_{j}$.
\item \textit{Aggregation} $A_{j,j}$: apply all aggregate vectors
		$t^{i}_{j}$ from columns $i < j$, with $\ell_{j,i} \neq 0$,
        to column $j$.
\end{itemize}

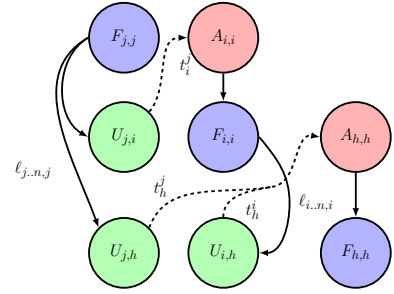
\begin{figure}
\centering
\begin{adjustbox}{width=.6\linewidth}
\begin{tikzpicture}
\scalefont{1.5}

\node[draw,circle,minimum size = 4em,ultra thick,fill=blue!30!white] (Fh) at (-1,1) {$F_{j,j}$};


\node[draw,circle,minimum size = 4em,ultra thick,fill=green!30!white] (Uhi) at (-1,-2) {$U_{j,i}$};
\node[draw,circle,minimum size = 4em,ultra thick,fill=green!30!white] (Uhj) at (-1,-5.5) {$U_{j,h}$};

\node[left] at (-3,-3) {$\ell_{j..n,j}$};
\draw[out=-150,in=150,-latex,ultra thick] (Fh.west) to (Uhi.west);
\draw[out=-150,in=110,-latex,ultra thick] (Fh.west) to (Uhj.north west);

\node[draw,circle,minimum size = 4em,ultra thick,fill=red!30!white] (Ai) at (2,1) {$A_{i,i}$};

\node[right] at (0.5,0.2) {$t^{j}_{i}$};
\draw[out=50,in=180,-latex,ultra thick,dashed]  (Uhi.north east) to (Ai.west);
\node[draw,circle,minimum size = 4em,ultra thick,fill=blue!30!white] (Fi) at (2,-2) {$F_{i,i}$};
\draw[-latex,ultra thick] (Ai) to (Fi);
\node[draw,circle,minimum size = 4em,ultra thick,fill=green!30!white] (Uij) at (2,-5.5) {$U_{i,h}$};

\node[right] at (4.1,-4) {$\ell_{i..n,i}$};
\draw[-latex,ultra thick] (Fi.east) to[out=-50,in=90] (4,-4) to[out=-90,in=0] (Uij.east);

\node[draw,circle,minimum size = 4em,ultra thick,fill=red!30!white] (Aj) at (6,-2) {$A_{h,h}$};
\node[right] at (2.5,-4.2) {$t^{i}_{h}$};
\node[left] at (0.5,-3.5) {$t^{j}_{h}$};

\draw[ultra thick,dashed]  (Uhj.north east) to[out=70,in=-170] (3.5,-3.5);
\draw[-latex,ultra thick,dashed]  (Uij.north) to[out=90,in=-170] (3.5,-3.5) to[out=10,in=170,] (Aj.west);

\node[draw,circle,minimum size = 4em,ultra thick,fill=blue!30!white] (Fj) at (6,-5.5) {$F_{h,h}$};
\draw[-latex,ultra thick] (Aj) to (Fj);

\end{tikzpicture}
\end{adjustbox}
\caption{\fanboth task dependencies for three columns
$j$, $i$ and $h$\label{fig.dependencies}}
\end{figure}

An example of dependencies among these tasks for three columns
$j$, $i$ and $h$, with $j < i$ and $j < h$, is depicted in Figure~\ref{fig.dependencies}.
After column $j$ has been factored, its updates to dependent columns
$i$ and $h$ can be computed. This corresponds to tasks $U_{j,i}$ and
$U_{j,h}$. Note that both these tasks require $\ell_{j:n,j}$, which
has to be \textit{fanned-out} to these two tasks. After these two tasks
have been processed, $t^{j}_{i}$ and $t^{j}_{h}$ have been computed.
$A_{i,i}$ can now be updated using $t^{j}_{i}$, after which $F_{i,i}$
is ready to be executed.
After that, the task $U_{i,h}$, which produces
$t^{i}_{h}$, can be executed. The two aggregate vectors $t^{j}_{h}$ and $t^{i}_{h}$
are then applied on column $h$
during the execution of task $A_{h,h}$,
requiring aggregate vectors to be \textit{fanned-in}. Finally, task
$F_{h,h}$ can be processed.
As can be observed, \fanboth indeed involves data exchanges that can
be observed in either \fanin or \fanout.

\subsection{Parallel algorithm and computation maps}

We now describe \fanboth in a parallel setting.
We assume a parallel distributed memory platform with $P$ processors
ranging from $\proc{1}$ to $\proc{P}$.
We assume that $\M{A}$ and $\M{L}$ are cyclically distributed
by supernodes of various sizes in a 1D way, as depicted in 
Figure~\ref{fig.cholesky_factor}. \REV{The maximum supernode size is limited to 150 columns.} This has the benefit of allowing a 
good load balancing of nonzero entries and computation per processor, 
although communication might not achieve optimal load balance.
An example of such a distribution using 4 distributed memory nodes, or processors, is depicted in Figure~\ref{fig.cholesky_factor}.

Ashcraft~\cite{ashcraft-dissertation} introduces the concept of computation
maps to guide the mapping of tasks onto processors. A mapping $\mathcal{M}$
is a two-dimensional grid that ``extends'' to the matrix size (i.e., $n$-by-$n$). Values represent node ranks computed using a closed-form generator expression. Therefore, the $n$-by-$n$ grid is not explicitly stored.
A mapping is said to be $1$-by-$P$ when
1 rank is found in each column of $\mathcal{M}$, $P$-by-1 when $P$ distinct values are
found in every column and $\sqrt{P}$-by-$\sqrt{P}$ when $\sqrt{P}$ distinct ranks are found on each row and column. 

A computation map $\mathcal{M}$ is used to map the tasks as follows:
\begin{itemize}
\item Tasks $A_{i,i}$ and $F_{i,i}$ are mapped onto node $\proc{\mathcal{M}_{i,i}}$
\item Tasks $U_{i,j}$ is mapped onto node $\proc{\mathcal{M}_{j,i}}$
\end{itemize}

In a parallel setting, aggregate vectors can be further accumulated on each
node to reflect the updates of all local columns residing on a given node $\proc{i}$
to a given column $j$. We let $a^{(\proc{i})}_{j}$ be such an aggregate vector. We have:
\[
a^{(\proc{i})}_{j} = \sum_{\begin{array}{c}\forall i<n~\textnormal{on}~\proc{i}\\
i ~\textnormal{updates}~ j\end{array}} t^{i}_{j}.
\]

Given a task mapping strategy $\mathcal{M}$, it is important to note that the factor columns 
$\ell_{i:n,i}$, $\forall i$, $1 \leq i \leq n$ need to be sent to at most the number
of distinct node ranks present in the lower triangular part of column $i$ of $\mathcal{M}$. Aggregate vectors need to be sent to the number of distinct ranks
in the lower triangular part of row $i$ of $\mathcal{M}$.

\begin{figure}[htbp]
\captionsetup[subfigure]{justification=centering}
\centering
\subfloat[][\Fanin\\\begin{adjustbox}{scale=0.65}$\mathcal{M}_{i,j} = mod(i,P)$\end{adjustbox}]{
\begin{adjustbox}{width=.3\linewidth}
\begin{tikzpicture}[cell/.style={rectangle,draw,minimum size=1cm}]
  \pgfmathtruncatemacro{\np}{4}
  \pgfmathtruncatemacro{\size}{6-1}
  \begin{scope}[yscale=-1]
  \foreach \x in {0,...,\size}
    \foreach \y in {0,...,\size} 
    {
       \pgfmathtruncatemacro{\map}{Mod(\x,\np)}
       \node [cell]  (\x\y) at (\x,\y) {\map };
    }
    \end{scope}
\end{tikzpicture}
\end{adjustbox}
}
\subfloat[][\Fanout\\\begin{adjustbox}{scale=0.65}$\mathcal{M}_{i,j} = mod(j,P)$\end{adjustbox}]{
\begin{adjustbox}{width=.3\linewidth}
\begin{tikzpicture}[cell/.style={rectangle,draw,minimum size=1cm}]
  \pgfmathtruncatemacro{\np}{4}
  \pgfmathtruncatemacro{\size}{6-1}
  \begin{scope}[yscale=-1]
  \foreach \x in {0,...,\size}
    \foreach \y in {0,...,\size} 
    {
       \pgfmathtruncatemacro{\map}{Mod(\y,\np)}
       \node [cell]  (\x\y) at (\x,\y) {\map };
    }
    \end{scope}
\end{tikzpicture}
\end{adjustbox}
}
\subfloat[][\Fanboth\\\begin{adjustbox}{scale=0.65}$\mathcal{M}_{i,j} = \begin{array}{l}mod(\min(i,j),P) +\\P\lfloor mod(\max(i,j),P)/P\rfloor\end{array}$\end{adjustbox}]{
\begin{adjustbox}{width=.3\linewidth}
\begin{tikzpicture}[cell/.style={rectangle,draw,minimum size=1cm}]
  \pgfmathtruncatemacro{\np}{4}
  \pgfmathtruncatemacro{\nr}{2}
  \pgfmathtruncatemacro{\size}{6-1}
  \begin{scope}[yscale=-1]
  \foreach \x in {0,...,\size}
    \foreach \y in {0,...,\size} 
    {    
      \pgfmathtruncatemacro{\tmp}{ mod(\x,\np)/\nr }
      \pgfmathparse{
        mod(\y,\nr) 
        +
        \nr*(\tmp)
                    }\let\map\pgfmathresult
      \pgfmathtruncatemacro{\map}{\map}
       \node [cell]  (\x\y) at (\x,\y) {\map };
    }     
    \end{scope}
\end{tikzpicture}
\end{adjustbox}
}
\caption{Three different computation maps, corresponding to algorithms in \fanin, \fanout and \fanboth\label{fig.mapping}}
\end{figure}
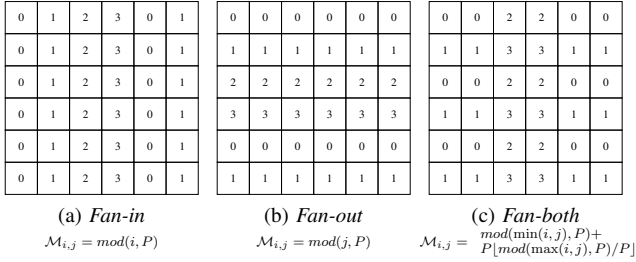

In~\cite{ashcraft-dissertation}, the author discusses the worst case
communication volume depending on which computation map is used.
The $\sqrt{P}$-by-$\sqrt{P}$ maps involve at most $\sqrt{P}$ nodes in each
communication step, while each step involves at most $P$ nodes for
either $P$-by-1 or 1-by-$P$ computation maps. The volume
is directly impacted by the number of nodes participating to each
communication step. However, the $\sqrt{P}$-by-$\sqrt{P}$ maps require
two kinds of messages to be exchanged (i.e. both factors and aggregate vectors) while 1-by-$P$ or $P$-by-1 only require one type of message. The 
latency cost is therefore higher for $\sqrt{P}$-by-$\sqrt{P}$ computation
maps.

Various formulations of the Cholesky factorization can generally be described by a \fanboth algorithm with appropriate computation maps.
For instance, \fanin and \fanout are good examples, as illustrated by Figure~\ref{fig.mapping}.
Our \sympack also uses these computation maps but it is not restricted
to such task assignments. This flexible design will allow us to 
derive and evaluate a wider range of task mapping and scheduling 
strategies.

\subsection{Impact of communication strategy}

Without loss of generality, parallel distributed memory algorithms
perform communication following two strategies. A data transfer
happening between two processors (or nodes) \psource and \ptarget can
be performed the following ways:
\begin{compactitem}
  \item \psource sends the data to \ptarget as soon as the data is available using a \emph{push} strategy,
  \item \ptarget gets the data from \psource as soon as data is required using a \emph{pull} strategy.
\end{compactitem}
These two strategies will be explained in detail below.

Another very important characteristic of communication protocols is whether a communication primitive
is two-sided or one-sided. The former requires \psource to issue a \textit{send} operation and \ptarget
to issue a matching \textit{receive} operation. Such strategy is employed in most MPI applications.
The latter strategy can be employed in two ways, and relies on the fact that the communication library is
able to write/read to/from a remote memory location. Either \psource \textit{puts} directly the data into
\ptarget's memory, or \ptarget \textit{gets} the data directly from \psource's memory. This type of communication
have been introduced in MPI-2 and refined in MPI-3, and is also available in other libraries such as \gasnet.

In the following, we discuss those strategies and their corresponding implications in
the context of the sparse Cholesky factorization, and more generally in the context of direct sparse
solvers.
Two kinds of messages can be exchanged 
throughout the factorization:
\emph{factors} and \emph{aggregate vectors}. The first type of messages 
corresponds to the entries in a column after it has been factorized,
or in other words, to a portion of the output data of the algorithm. The 
second type of messages is a temporary buffer
in which a given \psource will accumulate all its updates to a remote target 
column residing on \ptarget.

In the next two paragraphs, we suppose that two tasks $T_{i,j}$ and
$T_{j,j}$ are respectively mapped onto two processors \psource and
\ptarget. Task $T_{j,j}$ depends on data produced by task $T_{i,j}$.
Let
$M_{i,j}$ denote that data.

\paragraph{Push strategy}
First, \psource computes task $T_{i,j}$. As soon as it is done processing that 
task, it sends $M_{i,j}$ to processor \ptarget.
When processor \ptarget selects task $T_{j,j}$ to be executed, the first thing done is to post a \emph{receive} request, and wait until $M_{i,j}$ has been fetched. Once it is received,
task $T_{j,j}$ can be processed.

\paragraph{Pull strategy}
Let us now consider the \emph{pull} strategy. 
Processor \psource processes task $T_{i,j}$ and produces $M_{i,j}$.
Later on, processor \ptarget selects task $T_{j,j}$. It first sends a
message to processor \psource, requesting $M_{i,j}$ to be sent to
processor \ptarget. Once this transfer is completed, task $T_{j,j}$ can
be processed.

The key difference between the \emph{push strategy} and 
the \emph{pull strategy} is therefore which processor has
the responsibility to initiate the data exchange.

\subsection{Asynchronous communications and deadlock situations\label{sec.deadlocks}}

In the following, we discuss the use of asynchronous communication primitives in sparse
matrix solvers, and more specifically in the case of Cholesky factorization. Asynchronous
communications, or non-blocking communications, are often used in parallel applications
in order to achieve good strong scaling and deliver high performance.

In some situations though, asynchronous communications must be used with care. Communication
libraries have to resort to a certain number of buffers to perform multiple asynchronous communications
concurrently. However, the space for these buffers is a limited resource, and a communication library will certainly
run out of buffer space if too many asynchronous communications are performed concurrently. In such a case,
the communication primitives become blocking and deadlock might occur.
This latter case corresponds to the situation in which each processor
has only one send buffer and one receive buffer.

Let us consider a task graph and more precisely the simpler case where
that task graph is a directed tree, and analyze it in the context of matrix computation.
Let us also suppose that operations are performed on entire columns of an input matrix. Let us denote:
\begin{itemize}
\item $\M{A}$ an input matrix of dimension $n$ on which operations are going
		to be applied to columns.
\item $\mathcal{C}$ the set of columns in $\M{A}$ and distributed onto $P$
processors in a cyclic way. We have $\forall i \in \mathcal{C}, i\leq n$.
\item $\mathcal{T} = (V,E)$: a task tree, where $V$ is the set of vertices in
the tree and $E$ the set of directed edges between vertices in $E$.
\item $T_{i,j} \in V$ a task from a \emph{source} column 
		$i \in \mathcal{C}$ to a \emph{target} column
        $j \in \mathcal{C}, i \leq j$, of matrix $\M{A}$. Let us assume
        that column $j$ is modified after $T_{i,j}$ has been processed.
\item $M_{i,j} = (T_{i,j},T_{j,k}) \in E $ a dependency between two 
	tasks $T_{i,j}$ and $T_{j,k} \in V$. It corresponds to a communication
    if tasks are not mapped onto the same processor.
\end{itemize}

For sparse Cholesky factorization, such a task tree can be derived from the
\textit{elimination tree} of matrix $\M{A}$. We suppose that this elimination
tree has been labeled in a post-order fashion, which is generally the case.
Therefore, every edge $(T_{i,j},T_{j,k})$ in $E$ has to respect the 
constraint $i\leq j \leq k$. 

Suppose that each processor has only one send buffer and one
receive buffer, and that processors will push, or send, the newly
produced data as soon as it has been produced. Let us also assume
that prior to executing a task, the incoming data has to be received.

\begin{figure}[htbp]
\centering
\begin{adjustbox}{width=.85\linewidth}
\begin{tikzpicture}
\tikzstyle{every node}=[font=\scalefont{0.8}]

\colorlet{colp1}{blue!10}
\colorlet{colp2}{green!50!black!30!white}
\colorlet{colp3}{red!70!black!20!white}

\node[ellipse,minimum size=3em,fill=colp2] (3pp2) at (-2em,-4em) {$T_{3 P + 2,3 P + 2}$};
\node[ellipse,minimum size=3em,fill=colp1] (3pp1) at (8 em,-5em) {$T_{3 P+1,3 P + 2}$};

\begin{scope}[xshift=-2em]
\node[ellipse,minimum size=3em,fill=colp2] (pp2) at (-4 em,-10em) {$T_{P+2,3 P + 2}$};
\node[ellipse,minimum size=3em,draw=none,fill=none] (ppdots) at (0 em,-10em) {$\ldots$};
\node[ellipse,minimum size=3em,fill=colp3] (2p) at (3.5 em,-10em) {$T_{2 P,3 P + 2}$};
\begin{scope}[xshift=-3em]
\node[ellipse,minimum size=3em,fill=colp1] (pp1) at (-8 em,-10em) {$T_{P+1,3 P + 2}$};

\begin{scope}[xshift=3em]
\node[ellipse,minimum size=3em,fill=colp1] (1) at (-12 em,-16em) {$T_{1,P+1}$};
\node[ellipse,minimum size=3em,fill=none,draw=none] (dots) at (-4 em,-16em) {$\ldots$};
\node[ellipse,minimum size=3em,fill=colp2] (2) at (-7 em,-16em) {$T_{2,P+1}$};
\node[ellipse,minimum size=3em,fill=colp3] (p) at (-1 em,-16em) {$T_{P,P+1}$};
\end{scope}
\end{scope}
\end{scope}

\node[ellipse,minimum size=3em,fill=colp3] (3p) at (8 em,-9em) {$T_{3 P,3P +1}$};
\node[ellipse,minimum size=3em,fill=colp2] (2pp2) at (8 em,-13em) {$T_{2 P +2,2P+3}$};
\node[ellipse,minimum size=3em,fill=colp1] (2pp1) at (8 em,-18em) {$T_{2 P +1,2P+2}$};

\draw(3pp2) -- (pp1);
\draw(3pp2) -- (pp2);
\draw(3pp2) -- (2p);
\draw(3pp2) -- (3pp1);

\draw(pp1) -- (1);
\draw(pp1) -- (2);
\draw(pp1) -- (p);

\draw(3pp1) -- (3p);
\draw[dotted] (3p) -- (2pp2);
\draw(2pp2) -- (2pp1);

{\draw[ultra thick,green,transform canvas={shift={(+0.6 em, 0.0em)}},<-] (3pp2) -- (2p);}
{\draw[ultra thick,green,transform canvas={shift={(0.1 em, -0.5em)}},<-] (3pp2) -- (pp1);}
{\draw[ultra thick,green,transform canvas={shift={(+0.6 em, 0.0em)}},<-] (3pp2) -- (pp2);}

\draw[ultra thick,green,<-,transform canvas={shift={(0.3 em, 0.3em)}}] (pp1) -- (p) node[text=black,midway,xshift=1.6em,yshift=-0.1em] {1};
{\draw[red,ultra thick,transform canvas={shift={(-0.3 em, -0.3em)}},->,dashed] (pp1) -- (p);}
\draw[ultra thick,red,dashed,->,transform canvas={shift={(-0.3 em, -0.3em)}}] (pp1) -- (p) node[text=black,midway,xshift=-0.5em,yshift=-0.5em] {1};

\draw[ultra thick,green,<-,transform canvas={shift={(0.5 em, 0em)}}] (pp1) -- (2) node[text=black,midway,xshift=0.8em,yshift=-0.1em] {1};
\draw[ultra thick,red,dashed,->,transform canvas={shift={(-0.5 em, 0em)}}] (pp1) -- (2) node[text=black,midway,xshift=-0.8em,yshift=-0.1em] {1};

{\draw[ultra thick,red,transform canvas={shift={(-0.5 em, 0 em)}},->,dashed] (2pp2) -- (2pp1);}


\path let \p1=(3pp2), \p2 = (pp1) in node at ($(\x2 + 0.5*\x1 - 0.5*\x2,0.5*\y2 + 0.5*\y1) + (+1.3 em,-0.5 em)$) {$2$};
\path let \p1=(3pp2), \p2 = (pp2) in node at ($(\x2 + 0.5*\x1 - 0.5*\x2,0.5*\y2 + 0.5*\y1) + (+1.3 em,-0.5 em)$) {$2$};
\path let \p1=(3pp2), \p2 = (2p) in node at ($(\x2 + 0.5*\x1 - 0.5*\x2,0.5*\y2 + 0.5*\y1) + (+2 em,-0.5 em)$) {$2$};

\path let \p1=(2pp2), \p2 = (2pp1) in node at ($(\x1 + 0.5*\x1 - 0.5*\x2,0.5*\y2 + 0.5*\y1) + (-1.5 em,0)$) {$2$};


%
%
%
%
\end{tikzpicture}
\end{adjustbox}
\caption{A task tree where deadlock happens. Green solid arrows correspond to send operations (and their local order on each processor). Red dashed arrows correspond to receive operations (and their local order).}
\label{fig.deadlock}
\end{figure}
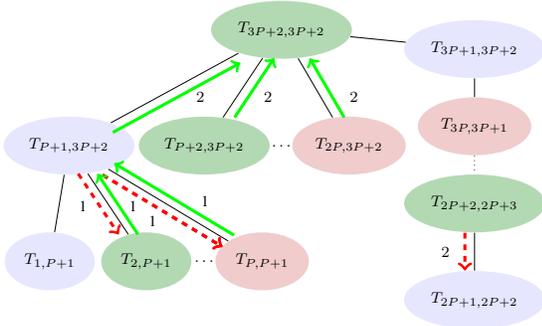

Such a task tree is depicted in 
Figure~\ref{fig.deadlock}, with different colors corresponding to distinct
processors. Let us describe how such a tree is processed by the $P$ processors.

First, each of the $P$ processors executes one task $T_{i,P+1}, 1\leq i \leq P$,
of the bottom level. Processors $\proc{2}$ to $\proc{P}$ send their respective data to 
processor $\proc{1}$, which receives each message one by one in a sequential way.

All processors can now compute tasks $T_{P+i,3P+2}, 1 \leq i \leq P$, and 
then send their respective data to processor $\proc{2}$ on which task
$T_{3P+2,3P+2}$ has been assigned. This consumes the send buffer of all processors
but processor $\proc{2}$. 

Processor $\proc{1}$ then computes task $T_{2P+1,2P+2}$ in the rightmost branch of
the tree. It cannot send the data because the send buffer is currently occupied.
Processor $\proc{2}$ is waiting for the incoming data to task $T_{2P+2,2P+3}$,
which cannot be sent by $\proc{1}$. Hence a deadlock situation.

In order to avoid this kind of situation, tasks and messages can be
scheduled in the following way:
\begin{itemize}
\item Process tasks $T_{i,j} \in V$ in non-decreasing order of target 
	column $j$, then in non-decreasing order of source column $i$.
\item Send message $M_{i,j} \in E$ in non-decreasing order of target 
	column $j$, then in non-decreasing order of source column $i$ and
    only if $M_{i,j} < T_{i',j'}$ with respect to this ordering where
    $T_{i',j'}$ is the next task scheduled onto this processor.
\end{itemize}

This problem has also been observed in~\cite{sid2014scaling} in the context
of multifrontal factorization, in which a similar criterion has to be used.


\section{Asynchronous task-based formulation\label{sec.dyn_scheduling}}\label{asynchronous}

\begin{figure*}[tbhp]
\centering
\begin{adjustbox}{width=.65\linewidth}
\includegraphics{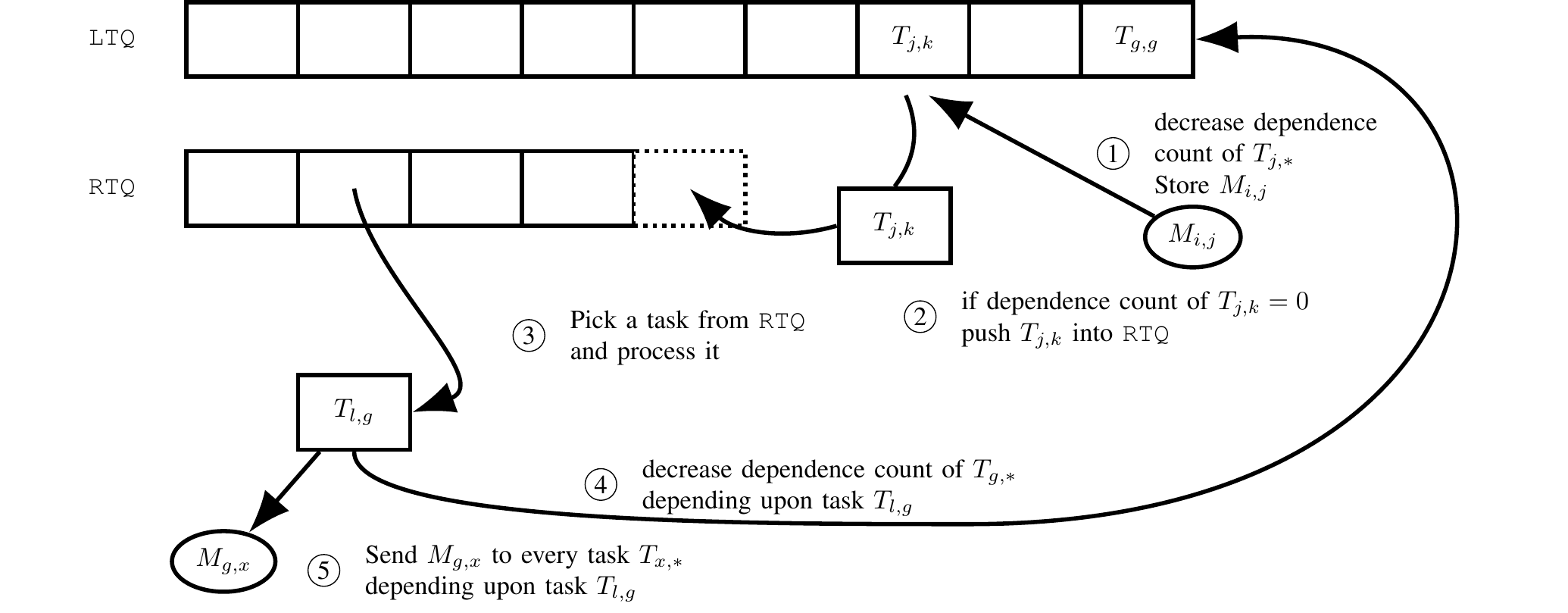}
\end{adjustbox}
\caption{Task scheduling in \sympack. Update of task dependence
in \ltq. Dynamic scheduling of local tasks in \rtq.}\label{task_schedule}
\end{figure*}

Modern platforms can be subject to high performance variability, and it is hard
to derive an accurate model of such platforms. Obtaining good static
scheduling strategies is therefore difficult.
For scheduling purposes, an application is often modeled using Directed Acyclic Graphs (or DAGs).
Computations are modeled as tasks, and represented by vertices in the graph while dependencies between
tasks are represented by the edges.
In the context of sparse matrix computations, there is no fixed task graph for a particular numerical kernel. 
The task graph is inherently depending upon the structure of the sparse
matrix on which the computation is going to be performed. This makes the use of advanced static scheduling techniques even harder to apply.

This motivated us to rely on a dynamic 
scheduling approach instead, which is by nature more amenable to cope with performance variations
and incomplete knowledge of the task graph. \REV{Even when the task graph is known, which is the case in sparse matrix factorizations,
task completions are hard to predict on a parallel platform and dynamic scheduling is an efficient way to deal with this issue.}
We propose the following data structure, where each processor has:
\begin{itemize}
  \item a \textit{local task queue} (\ltq), containing all the tasks statically mapped onto this processor and awaiting execution,
  \item a \textit{ready task queue} (\rtq), containing all the tasks for which precedence constraints have been satisfied and 
    that can therefore be processed.
\end{itemize}
This is illustrated in Figure~\ref{task_schedule}.

A task $T_{s,t}$ is represented by a \textit{source} supernode $s$ 
and a \textit{target} supernode $t$ on which computations have to be
applied. Each task also has an incoming dependency counter, initially
set to the number of incoming edges.

As \sympack implements a factorization similar to \fanboth, three types 
of tasks have to be dealt with. 
Similarly, a message $M_{s,t}$ exchanged to satisfy the dependence 
between tasks mapped onto distinct processors is labeled by the
\textit{source} supernode $s$ of the receiving task and the
\textit{target} supernode $t$ of the receiving task.

The overall mechanism that we propose is the following: whenever a task is completed,
processors owning dependent tasks are notified that new input data is now available.

As soon as a processor is done with its current computation, it periodically handles
these incoming notifications by issuing one-sided \textit{gets}.
This \textit{get} operation can either be a non-blocking 
communication or a blocking communication. The incoming dependency counter of the 
corresponding task is decremented when the communication has been completed.
This corresponds to a strategy similar to the \emph{pull} strategy discussed earlier, the only difference being that \psource directly
notifies \ptarget rather than \ptarget periodically requesting data
to \psource. 

When a task from the \ltq has all its dependencies satisfied (i.e., when its
dependency counter reaches zero) then it is moved to the \rtq, and is now 
ready for execution.
The processor then picks a task from the \rtq and executes it. If multiple tasks are 
available in the \rtq, then the next task that will be processed
is picked according to a dynamic scheduling policy. As a first step,
we use the same criterion for picking a task in the \rtq than the
criterion that prevents deadlocks. Evaluating different scheduling 
policies will be the subject of a future analysis.



\subsection{Data-driven asynchronous communication model\label{sec.comm_protocol}}

Communications are becoming a bottleneck in most scientific computing applications. This is even more true for sparse linear algebra kernels, which often exhibit
a higher communication to computation ratio. Moreover, modern parallel platforms
have to exploit interconnects that are more complex than in the past, and often
display a deeper hierarchical structure. 
Larger scale also has a side-effect which can be observed on most modern platforms: performance variability. 
In the following, we propose a communication protocol for our parallel sparse Cholesky implementation that allow the communications to drive the scheduling.
This is a crucial piece toward an Asynchronous Task execution model.

We use the \upcxx PGAS library~\cite{ZKDSY14, upcxx-web} for communicating between distributed memory nodes.
\upcxx is built on top of \gasnet~\cite{gasnet-web}, and introduces
several parallel programming features useful for our implementation. 

First, it provides \textit{global pointers} for accessing memory
locations on remote nodes. Using the \textit{get} and \textit{put} functions,
one can transfer the data between two nodes in a one-sided way. Moreover, these
transfers are handled by RDMA calls, and are therefore generally performed
without interrupting the remote processor. Using this concept of global pointers, \upcxx also allows us to allocate and deallocate memory on a node from a remote node.

Another useful feature is the ability to perform \textit{remote asynchronous function calls}.
A processor can submit a function for execution on a remote node. It gets pushed into a queue
that the remote processor executes when calling the \upcxx \textit{progress} function.

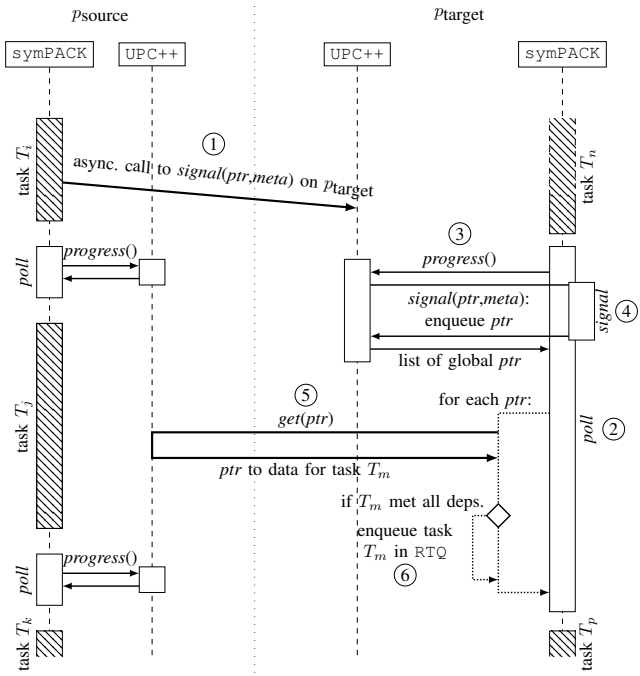
\begin{figure}[htbp]
\centering
\begin{adjustbox}{width=\linewidth}
\begin{tikzpicture}

\begin{scope}[yscale=-1]
\node (ps) at (1,-1.5) {\psource};
\node[rectangle,draw=black,fill=white] (sps) at (0,-0.75) {\sympack};
\node[rectangle,draw=black,fill=white] (upcxxs) at (2,-0.75) {\upcxx};
\path[draw=black,dashed] (upcxxs.south) -- (2,11);
\path[draw=black,dashed] (sps.south) -- (0,11);

\path[fill=white] ($ (0,0) + (-0.25,0.5)$) rectangle +(0.5,2);
\path[draw=black,pattern=north west lines, pattern color=black] ($ (0,0) + (-0.25,0.5)$) rectangle +(0.5,2);
\node[rotate=90] at (-0.5, 1.5) {task $T_i$};

\path[very thick,draw=black,-latex] (0.25,1.75) -- (6,2.25);
\node[rotate=-5] at (3.4, 1.65) {async. call to \textit{signal}(\textit{ptr},\textit{meta}) on \ptarget};
\node at (3.2, 1.0) {\circled{1}\label{fig.notify.step1}};
\path[draw=black,fill=white] ($ (0,0) + (-0.25,3)$) rectangle +(0.5,1);
\node[rotate=90] at (-0.5, 3.5) {$\textit{poll}$};
\path[draw=black,fill=white] ($ (2,0) + (-0.25,3.25)$) rectangle +(0.5,0.5);
\path[thick,draw=black,-latex] (0.25,3.375) -- (1.75,3.375);
\node at (1, 3.1) {\textit{progress}()};
\path[thick,draw=black,-latex] (1.75,3.625) -- (0.25,3.625);

\path[draw=black,fill=white] ($ (0,0) + (-0.25,4.5)$) rectangle +(0.5,4);
\path[draw=black,,pattern=north west lines, pattern color=black] ($ (0,0) + (-0.25,4.5)$) rectangle +(0.5,4);
\node[rotate=90] at (-0.5, 6.5) {task $T_j$};

\path[draw=black,fill=white] ($ (0,0) + (-0.25,9)$) rectangle +(0.5,1);
\node[rotate=90] at (-0.5, 9.5) {$\textit{poll}$};
\path[draw=black,fill=white] ($ (2,0) + (-0.25,9.25)$) rectangle +(0.5,0.5);
\path[thick,draw=black,-latex] (0.25,9.375) -- (1.75,9.375);
\node at (1, 9.1) {\textit{progress}()};
\path[thick,draw=black,-latex] (1.75,9.625) -- (0.25,9.625);

\path[fill=white] ($ (0,0) + (-0.25,10.5)$) rectangle +(0.5,0.5);
\path[pattern=north west lines, pattern color=black] ($ (0,0) + (-0.25,10.5)$) rectangle +(0.5,0.5);
\path[draw=black] ($ (0,0) + (-0.25,11)$) -- +(0,-0.5) -- +(0.5,-0.5) -- +(0.5,0); 
\node[rotate=90] at (-0.5, 10.75) {task $T_k$};

\path[draw=black,loosely dotted] (4,-2) -- (4,11.5);

\node (pt) at (8,-1.5) {\ptarget};
\node[rectangle,draw=black,fill=white] (spt) at (10,-0.75) {\sympack};
\node[rectangle,draw=black,fill=white] (upcxxt) at (6,-0.75) {\upcxx};

\path[draw=black,dashed] (spt.south) -- (10,11);

\path[fill=white,draw=none] ($ (10,0) + (-0.25,0.5)$) rectangle +(0.5,2.25);
\path[draw=none,pattern=north west lines, pattern color=black] ($ (10,0) + (-0.25,0.5)$) rectangle +(0.5,2.25);
\path[draw=black] ($ (10,0) + (-0.25,0.5)$) +(0.5,0) edge[dashed] +(0.5,1.5) +(0.5,1.5) -- +(0.5,2.25) +(0.5,2.25) -- +(0,2.25) -- +(0,1.5) +(0,1.5) edge[dashed] +(0,0); 
\node[rotate=90] at (10.5, 1.625) {task $T_n$};

\path[draw=black,fill=white] ($ (10,0) + (-0.25,3)$) rectangle +(0.5,7.125);
\node[rotate=90] at (10.5,6.5625) {$\textit{poll}$};
\node at (11, 6.5625) {\circled{2}\label{fig.notify.step2}};

\path[thick,draw=black,-latex] (9.75,3.5) -- (6.25,3.5);
\node at (8, 3.25) {\textit{progress}()};
\node at (8, 2.75) {\circled{3}\label{fig.notify.step3}};

\path[draw=black,dashed] (upcxxt.south) -- (6,11);
\path[draw=black,fill=white] ($ (6,0) + (-0.25,3.25)$) rectangle +(0.5,2);

\path[thick,draw=black,-latex] (6.25,3.75) -- (10.375,3.75) |- (6.25,4.75);
\path[draw=black,fill=white] (10.125,3.7) rectangle +(0.5,1.125);
\node[left] at (9.7, 4.25) {\begin{tabular}{ c} \textit{signal}(\textit{ptr},\textit{meta}): \\enqueue \textit{ptr}\end{tabular}};
\node[rotate=90] at (10.8, 4.2625) {$\textit{signal}$};
\node at (11.25, 4.2625) {\circled{4}\label{fig.notify.step4}};

\path[thick,draw=black,-latex] (6.25,5) -- (9.75,5);
\node at (8, 5.25) {list of global \textit{ptr}};

\node at (8.5, 6) {for each \textit{ptr}:};
\path[thick,draw=black,-latex,densely dotted] (9.75,6.25) -- (8.75,6.25) |- (9.75,9.75);
\node at (5, 5.9) {\circled{5}\label{fig.notify.step5}};
\node at (5, 6.375) {\textit{get}(\textit{ptr})};
\path[very thick,draw=black,-latex] (8.75,6.625) -- (2,6.625) -- (2,7.125) -- (8.75,7.125);
\node at (5, 7.375) {\textit{ptr} to data for task $T_m$};

\node[left] at (8.6, 8) {if $T_m$ met all deps.};
\node[thick,draw=black,fill=white,diamond,minimum width=0.5,minimum height=0.5] (ifdone) at (8.75,8.25) {};
\path[thick,draw=black,-latex,densely dotted] (ifdone.west) -- (8.25,8.25) |- (8.75,9.5);
\node[left] at (8.2, 9) {\begin{tabular}{c} enqueue task\\ $T_m$ in \rtq \\ 
\circled{6}\label{fig.notify.step6}
\end{tabular}};

\path[fill=white,draw=none] ($ (10,0) + (-0.25,10.5)$) rectangle +(0.5,0.5);
\path[draw=none,pattern=north west lines, pattern color=black] ($ (10,0) + (-0.25,10.5)$) rectangle +(0.5,0.5);
\path[draw=black] ($ (10,0) + (-0.25,10.5)$) -- +(0.5,0) -- +(0.5,0.5) +(0.5,2) +(0,2) +(0,0.5) -- +(0,0); 

\node[rotate=90] at (10.5, 10.75) {task $T_p$};

\end{scope}
\end{tikzpicture}
\end{adjustbox}
\caption{Data exchange protocol in \sympack. Notifications are performed using \upcxx asynchronous tasks, actual data is fetched with one-sided get.}
\label{fig.notify_process}
\end{figure}

We consider a data notification and communication process which is heavily based on
these two features of \upcxx. This process is depicted in Figure~\ref{fig.notify_process}.
Let us suppose that at the end of a computation, \psource has produced some data that needs to be sent out to \ptarget. 

First, \psource notifies \ptarget that some data has been produced by sending it a pointer \textit{ptr} to the data along with some meta-data \textit{meta}. This is done by doing an asynchronous function call to a {\textit{signal}(\textit{ptr},\textit{meta})} function on \ptarget directly from \psource, and is referred to as step 1 in the diagram.

When \ptarget finishes its current computation, it calls a \textit{poll} function (step~2), whose main role is to watch for incoming 
communications and do the book-keeping of task dependencies. This function resorts
to \upcxx \textit{progress} function to execute all asynchronous calls to the
\textit{signal}(\textit{ptr},\textit{meta}) function, which enqueues \textit{ptr}
and \textit{meta} into a list. This corresponds to steps~3 and~4.
The next step in the \textit{poll} function is to go through that list of global \textit{ptr} and issue a \textit{get} operation to pull the data (step~5). Note that this \textit{get} can be asynchronous, but for the sake of simplicity, we suppose a blocking \textit{get} operation here.
Once the \textit{get} operation is completed, the \textit{poll} function updates
the dependencies of the every task $T_m$ that will be using this data (which can be found by looking at the meta-data \textit{meta}). If all dependencies of a task are met, that task is moved into 
the list of ready-tasks \rtq (at step~6).

Finally, \ptarget resumes its work by selecting a task from \rtq and run it.

As mentioned before, two types of data are encountered in Cholesky factorization: 
factors and aggregate vectors. Factors represent the output of the algorithm.
The procedure described in Figure~\ref{fig.notify_process} can be applied to
these factors in a straightforward way.
Aggregate vectors, however, are temporary data. Hence, they need to be deleted when
not required anymore; that is after step~5 of the process has been completed. \upcxx allows a process to deallocate memory on a remote process using a global pointer to that memory zone. Therefore, when dealing with aggregate vectors, \ptarget will deallocate the data pointed by \textit{ptr} on \psource after it is done fetching it, without interrupting \psource.




\section{Performance evaluation}\label{performance}

In this section, we present the performance of the sparse
Cholesky factorization implemented in our solver \sympack.
Our experiments are conducted on the NERSC Edison supercomputer, which is based
on Cray XC30 nodes. Each node has two Intel(R) Xeon(R) CPU E5-2695 v2 ``Ivy bridge'' 
processors with 12 cores running at 2.40GHz and 64GB of memory~\cite{edison}.

We evaluate the performance of \sympack, our parallel asynchronous task-based sparse
Cholesky implementation using a set of matrices from the University of Florida
Sparse Matrix Collection~\cite{FloridaMatrix}. A description of each
matrix can be found in Table~\ref{tab.matrices}.

In this paper, we analyze the performance of \sympack in a 
distributed memory setting only. Therefore, all experiments
are conducted without multi-threading (which is commonly
referred to as ``flat-MPI'').

For sparse Cholesky factorization, the amount of fill-in that occurs depends on where the nonzero elements are in the matrix.  Permuting (or ordering) the matrix symmetrically changes its zero-nonzero structure, and hence changes the amount of fill-in in the factorization process.
In our experiments, a fill-reducing ordering computed using 
\scotch~\cite{scotch} is applied to the original matrix. The
\scotch library contains an implementation of the nested dissection algorithm~\cite{GeorgeND}
to compute a permutation that reduces the number
of fill-in entries in the Cholesky factor.

\begin{table}[htbp]
\centering
\begin{adjustbox}{width=\linewidth}
\begin{tabular}{| c | l | c | c | c | }
\hline
\multicolumn{5}{| c |}{Matrices from UFL sparse matrix collection}\\
\hline
Name & Type & $n$ & $nnz(A)$ & $nnz(L)$\\
\hline
{boneS10} & \begin{tabular}{l}3D trabecular bone
\end{tabular} & 914,898  & 20,896,803 & 318,019,434 \\
{bone010} & \begin{tabular}{l}3D trabecular bone
\end{tabular} & 986,703  & 24,419,243  & 1,240,987,782\\
{G3\_circuit} & \begin{tabular}{l}Circuit simulation problem\end{tabular} & 1,585,478 & 4,623,152 & 107,274,665 \\
{audikw\_1} & \begin{tabular}{l} Symmetric rb matrix \end{tabular} & 943,695 & 39,297,771 & 1,221,674,796 \\
{af\_shell7} & \begin{tabular}{l} Sheet metal forming,\\ positive definite\end{tabular} & 504,855 & 9,042,005 & 104,329,190\\
{Flan\_1565} & \begin{tabular}{l}3D model of a steel
flange,\\hexahedral finite elements\end{tabular} & 1,564,794 & 57,865,083 & 1,574,541,576\\
\hline
\end{tabular}
\end{adjustbox}
\caption{Characteristics of matrices used in the experiments\label{tab.matrices}}
\end{table}

\subsection{Impact of communication and scheduling strategy}

\begin{figure}[htbp]
\centering
\begin{adjustbox}{width=\linewidth}
\input{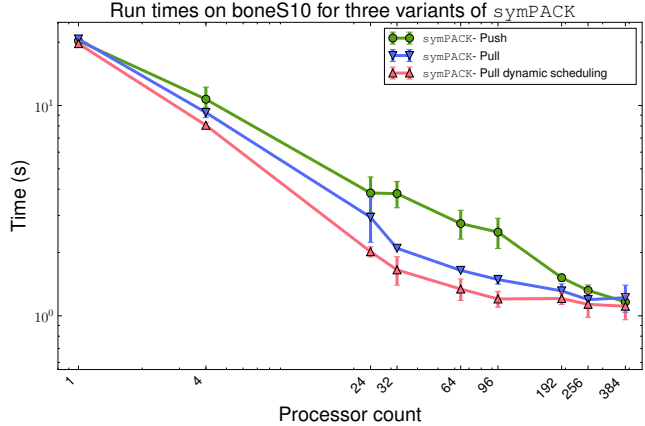}
\end{adjustbox}
\caption{ Impact of communication strategy and scheduling on \sympack performance\label{fig.comm_impact}}
\end{figure}

First, we aim at characterizing the impact of the communication
strategy used during Cholesky factorization. We also aim at
evaluating the impact of the dynamic scheduling described
in Section ~\ref{sec.dyn_scheduling}. To this end, we conduct
a strong scaling experiment using the \textit{boneS10} matrix from 
the University of Florida Sparse Matrix collection~\cite{FloridaMatrix}. 
Run times are averaged out of two runs.
Results are depicted in Figure~\ref{fig.comm_impact}, with error
bars representing standard deviations.
In this experiment, three variants of \sympack are compared: 
\textit{Push}, \textit{Pull} and \textit{Pull + dynamic scheduling}.

The \textit{Push} variant of \sympack is based on a two-sided push
communication protocol implemented using MPI. It uses the scheduling 
constraints introduced in Section~\ref{sec.deadlocks} to prevent
deadlocks. These constraints apply to both computations and 
communications.

The \textit{Pull} variant implements a one-sided pull 
communication protocol using \upcxx, but relieves the constraints on 
communications while still respecting the constraints on
computations. As a result, both \textit{Push} and \textit{Pull} 
executes the same static schedule for computations, but organize
communication in two different ways.

We observe in Figure~\ref{fig.comm_impact} that the \textit{Pull}
variant of \sympack outperforms
the \textit{Push} variant. This confirms that the communication protocol described in Section~\ref{sec.comm_protocol} and that
relies on \upcxx to perform the one-sided communications displays
a negligible overhead compared to a two-sided communication strategy
using MPI.

This performance difference confirms that the sorting
criterion that needs to be applied on both tasks and outgoing 
communications when using a push strategy also significantly
constrains the schedule. Removing the
constraints on how communications are scheduled while avoiding
still deadlocks through the use of the \textit{Pull} strategy allows
to achieve a better scalability.

This trend is further improved by using a dynamic scheduling policy 
in conjunction with the \textit{Pull} strategy. This confirms the
dynamic scheduling as described in Section~\ref{sec.dyn_scheduling}
is a good way to improve scalability in the context of sparse
matrix computations. In the rest of the paper, results 
corresponding to \sympack will correspond to the 
\textit{Pull + dynamic scheduling} variant.

\subsection{Strong scaling}

In the next set of experiments, we evaluate the strong scaling of our
sparse symmetric solver \sympack. We compare its performance to two
state-of-the-art parallel symmetric solvers: \mumps 5.0~\cite{mumps} and \pastix 5.2.2~\cite{pastix}.
The package \mumps is a well-known sparse solver based on the multifrontal approach
and that implements a symmetric factorization.
The code \pastix is based on a right-looking supernodal formulation.


We also provide the run times achieved by \superlu 4.3~\cite{LiDemmel2003,LiSuperLU} as a reference. Note
that \superlu is not a symmetric code and therefore requires twice as much memory
and floating point operations (if the columns are factored in the \textit{same} order). However, it is well known for its good strong scaling.
Therefore, only scalability trend rather than run times should be compared.

\REV{The same ordering, \scotch, is used for all solvers presented
in the experiments.}
As this paper focuses solely on distributed memory platforms, 
neither \pastix, \mumps nor \superlu are using multi-threading.
Furthermore, the term \emph{processor} corresponds to a 
distributed memory process.
Each data point corresponds to the average of three runs.

\begin{figure}[htbp]
\centering
\begin{adjustbox}{width=\linewidth}
\input{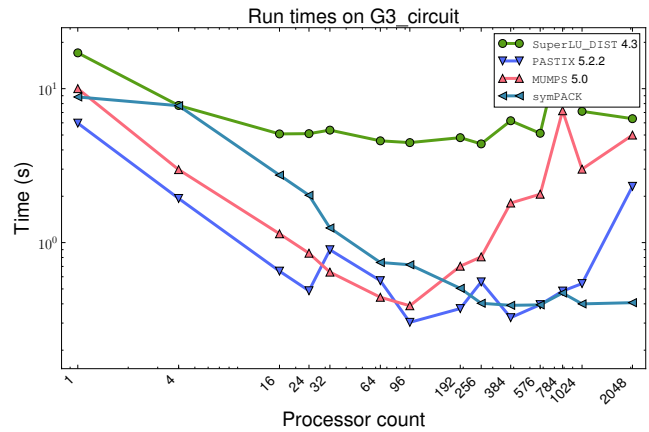}
\end{adjustbox}
\caption{Strong scaling of \sympack on G3\_circuit\label{fig.ss_g3}}
\end{figure}

\begin{figure}[htbp]
\centering
\begin{adjustbox}{width=\linewidth}
\input{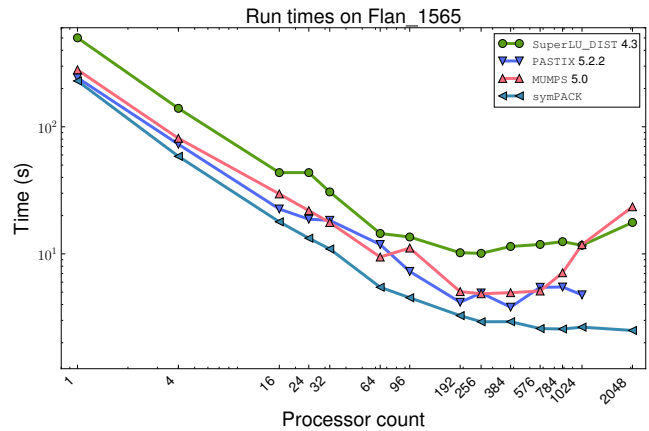}
\end{adjustbox}
\caption{Strong scaling of \sympack on Flan\_1565\label{fig.ss_Flan_1565}}
\end{figure}

On the G3\_circuit matrix, for which results are depicted
in Figure~\ref{fig.ss_g3}, \mumps and \pastix perform better
when using up to 96 and 192 processors respectively. On larger platform,
\sympack becomes faster than these two state-of-the-art solvers,
displaying a better strong scaling. 
The average speedup against the fastest solver for this specific 
matrix is 1.07, with a minimum value of 0.24 and a maximum value 
of 5.70 achieved when using 2048 processors.

The performance of \sympack on a smaller number of processors can 
be explained by the data structures which are used to reduce the memory
usage at the expense of more expensive indirect addressing operations. The G3\_circuit matrix being extremely sparse, it
is very likely that simpler structures with lower overhead would yield a higher level of performance.
In terms of scalability, \sympack displays a favorable trend when compared to \superlu, which 
scales up 192 processors on the expanded problem.

\begin{figure}[htbp]
\centering
\begin{adjustbox}{width=\linewidth}
\input{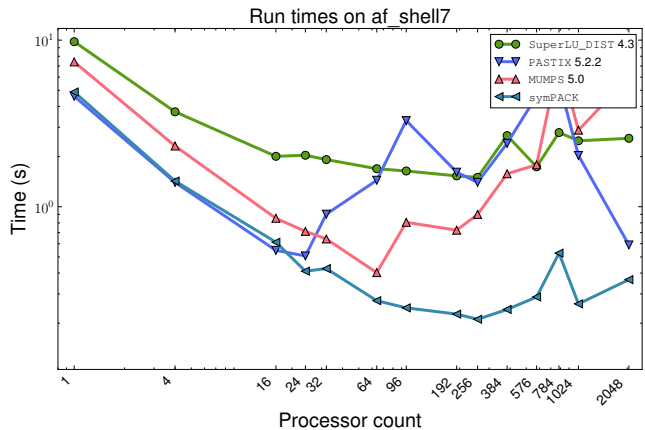}
\end{adjustbox}
\caption{Strong scaling of \sympack on af\_shell7\label{fig.ss_af_shell7}}
\end{figure}

\begin{figure}[htbp]
\centering
\begin{adjustbox}{width=\linewidth}
\input{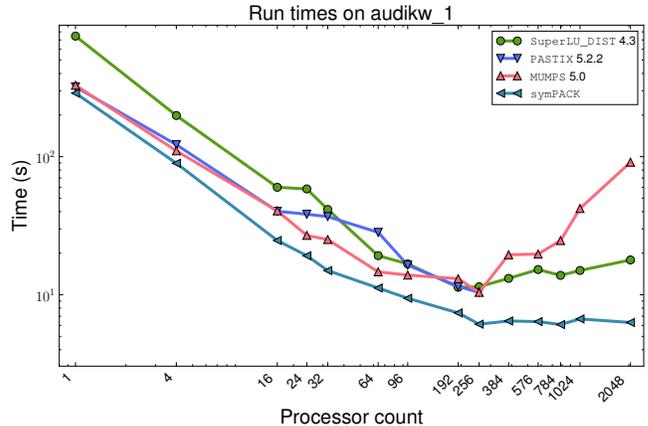}
\end{adjustbox}
\caption{Strong scaling of \sympack on audikw\_1\label{fig.ss_audikw_1}}
\end{figure}

\begin{figure}[htbp]
\centering
\begin{adjustbox}{width=\linewidth}
\input{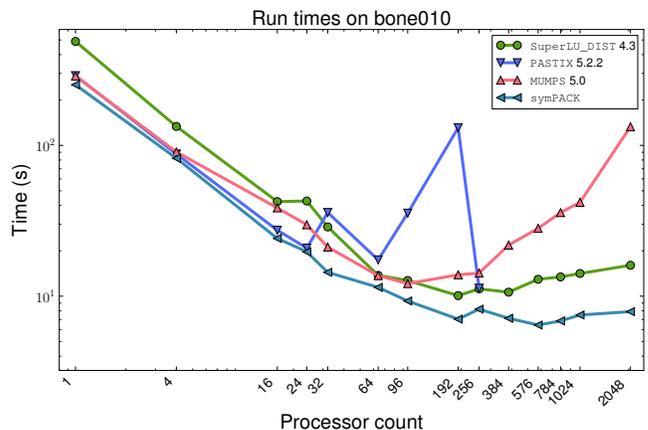}
\end{adjustbox}
\caption{Strong scaling of \sympack on bone010\label{fig.ss_bone010}}
\end{figure}

\begin{figure}[htbp]
\centering
\begin{adjustbox}{width=\linewidth}
\input{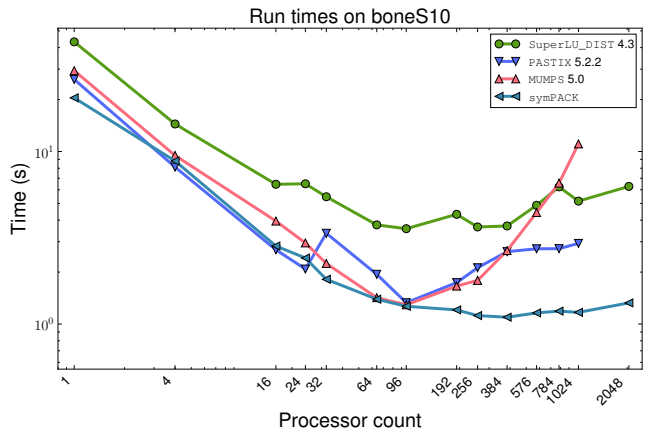}
\end{adjustbox}
\caption{Strong scaling of \sympack on boneS10\label{fig.ss_boneS10}}
\end{figure}

On other problems, \sympack is faster than all alternatives,
as observed on Figures~\ref{fig.ss_Flan_1565}, 
\ref{fig.ss_af_shell7}, \ref{fig.ss_audikw_1}, 
\ref{fig.ss_bone010}, and~\ref{fig.ss_boneS10}. Detailed speedups over the best symmetric solver and the best overall solver (thus including \superlu) are presented in Table~\ref{tab.speedup}.
The highest average speedup is achieved on the af\_shell7 problem,
for which \sympack can achieve an average speedup of 3.21 over the
best of every other solver. The corresponding minimum speedup is
0.89 while the maximum is 7.77.

\begin{table}
\centering
\begin{tabular}{| c | r | r| r | r | r | r |}
\cline{2-7}
\multicolumn{1}{ c |}{} & \multicolumn{3}{| c |}{Speedup vs. sym.} & \multicolumn{3}{| c |}{Speedup vs. best} \\
\hline
Problem & min & max & \textbf{avg.} & min & max & \textbf{avg.} \\
\hline
G3\_circuit  & 0.24  & 5.70  & \textbf{1.07} & 0.24  & 5.70 & \textbf{1.07} \\
\hline 
Flan\_1565   & 1.06  & 9.40  & \textbf{2.11} & 1.06  & 7.07 & \textbf{1.94} \\
\hline
af\_shell7   & 0.89  & 10.61 & \textbf{3.61} & 0.89  & 7.77 & \textbf{3.21} \\
\hline
audikw\_1    & 1.11  & 14.46 & \textbf{3.14} & 1.11  & 2.84 & \textbf{1.77} \\
\hline
boneS10      & 0.86  & N.A. & N.A. & 0.86  & 4.73 & \textbf{1.75} \\
\hline
bone010      & 1.06  & 16.83 & \textbf{3.34} & 1.06  & 2.03 & \textbf{1.47} \\
\hline
\end{tabular}
\caption{Speedup of \sympack over state-of-the-art solvers 
\label{tab.speedup}}
\end{table}

Interestingly, \superlu is the fastest of the state-of-the-art 
solvers on the audikw\_1 and bone010 matrices when using more 
than 384 processors. In those two cases, \sympack achieves an
average speedup of respectively 1.77 and 1.47. If the memory constraint
is such that one cannot run an unsymmetric solver like \superlu, 
then \sympack achieves an average speedup of respectively 3.14 
and 3.34 over the best symmetric solver.
Note that on the boneS10 matrix, neither \pastix nor \mumps succeeded using 2048 processors.

Altogether, the experiments confirmed that the asynchronous task
paradigm used in \sympack leads to promising practical
results in the context of sparse matrix computations. When used in
conjunction with a dynamic scheduling strategy, \sympack outperforms
the state-of-the-art symmetric solvers. This is crucial for memory
constrained environment. However, even when the amount of memory is 
sufficient to perform a $LU$ factorization instead of the Cholesky 
factorization, the approach proposed in this paper allows \sympack
to efficiently leverage the benefit of doing less computations, thus demonstrating the importance of symmetric solvers.

\section{Conclusion}\label{conclusion}

In this paper, we proposed a novel asynchronous task based approach
and studied it in the context of sparse matrix computations. For this
specific type of algorithms, whose performance is critical to numerous 
scientific applications, the communication strategy has to be chosen
carefully. We described a potential deadlock situation that can be
faced by any solver relying solely on asynchronous communications if
the communication library runs out of buffer space, and proposed a
scheduling constraint that allows these deadlock situations to be avoided.

The dynamic scheduling approach proposed in this paper successfully 
benefited the task formalism that we have described. The implementation
of these techniques was made significantly easier by relying on
new communication primitives and asynchronous function launch 
capabilities offered by \upcxx. Our numerical
experiments show that our solver \sympack significantly outperforms
state-of-the-art symmetric solvers on distributed memory platforms,
simultaneously demonstrating the validity of our approach and the
low-overhead and benefit of using new generation communication 
libraries such as \upcxx.

Leveraging the ever larger number of cores within a shared memory 
node to efficiently exploit the available concurrency offered by an
asynchronous task model coupled with a dynamic scheduling policy will 
be our immediate future work. Another important future work will be
to investigate how dynamic scheduling policies can be optimized in the
particular context of sparse linear algebra.

\section*{Acknowledgments}
This work was partially supported by the 
Scientific Discovery through Advanced Computing (SciDAC)
program funded by U.S. Department of Energy, Office of Science, 
Advanced Scientific Computing Research and Basic Energy Sciences 
(M. J. and E. N.), and the X-STACK program funded by U.S. Department of Energy, Office of Science, Advanced Scientific Computing Research
(Y.Z. and K.Y.).

\bibliographystyle{IEEEtran}
\bibliography{manuscript}

\end{document}